\documentclass[twocolumn,pre,floatfix]{revtex4}

\usepackage{epsfig}
\usepackage{amsmath}
\usepackage{times}

\begin{document}

\title{On coarse-graining by the Inverse Monte Carlo method: 
       Dissipative Particle Dynamics simulations made to 
       a precise tool in soft matter modeling}
  
\author{
Alexander P. Lyubartsev$^{[a]}$, 
Mikko Karttunen$^{[b]}$ , 
Ilpo Vattulainen$^{[c]}$,  and Aatto Laaksonen$^{[a]}$ }  

\address{
$^{[a]}$ Division of Physical Chemistry, Arrhenius Laboratory, 
Stockholm University, S--106~91, Stockholm, Sweden\\ 
$^{[b]}$ Biophysics and Statistical Mechanics Group, Laboratory of 
Computational Engineering,  
Helsinki University  of Technology, P.\,O.\,Box 9400,
FIN--02015 HUT Helsinki, Finland\\ 
\noindent
$^{[c]}$ Laboratory of Physics and Helsinki Institute of Physics,  
Helsinki University of Technology, P.\,O.\,Box 1100  
FIN--02015 HUT,  Helsinki, Finland}

\begin{abstract}

We present a promising coarse-graining strategy for 
linking micro- and mesoscales of soft matter systems. 
The approach is based on effective pairwise interaction 
potentials obtained from detailed atomistic molecular 
dynamics (MD) simulations, which are then used in 
coarse-grained dissipative particle dynamics (DPD) 
simulations. Here, the effective potentials were obtained
by applying the Inverse Monte Carlo method [Lyubartsev 
$\&$ Laaksonen, Phys. Rev. E. {\bf 52}, 3730 (1995)] 
on a chosen subset of degrees of freedom described in 
terms of radial distribution functions. In our first 
application of the method, the effective potentials 
were used in DPD simulations of aqueous NaCl solutions. 
With the same computational effort we were able to 
simulate systems of one order of magnitude larger as 
compared to the MD simulations. The results from the 
MD and DPD simulations are found to be in excellent 
agreement. 

\vspace*{12pt}
\end{abstract}

\maketitle

\section{Introduction}

Classical molecular dynamics computer simulations 
with site-site pair potentials can currently be run 
for molecular systems consisting of the order of $10^4$ 
atoms, corresponding typically to a system size of 
50\,--\,60\,\AA. Corresponding time scales using the 
same approach are typically of the order of 10\,§ns. 
This may be enough for simulations of isotropic liquids 
of simple molecules, while for many complex systems 
the requirements for the length and time scales are 
much more extensive.

The restrictions of the detailed all-atom description 
are especially severe in soft matter systems such as 
biomembranes and polyelectrolytes, although these are 
just two examples of the diversity of soft materials. 
This is mainly due to a wide range of time and length 
scales associated with these systems, and it is obvious 
that these problems become particularly pronounced in 
studies of dynamic processes which usually take place 
under hydrodynamic conditions. The primary aim in 
coarse-graining is to build a bridge between different 
time and length scales. Although there is no unique way 
to do this, the coarse-graining procedure always leads 
to a reduction of the number of degrees of freedom and 
thus alleviates the computational burden. The challenge 
is to establish systematic coarse-graining schemes that 
allow one to develop simplified model systems in terms 
of the information extracted from the underlying microscopic 
systems. Thus far, various approaches have been used to 
coarse-grain atomic and molecular systems. Below, we 
will briefly discuss some of them.

As a relatively generic approach, the dissipative particle 
dynamics (DPD) \cite{hoogerbrugge92a,koelman93a,espanol95a} 
has been used to study coarse-grained particle and polymeric 
systems, see e.g. Ref.~\cite{warren98a} for a review. In 
our approach we use DPD as a thermostating method which  
is discussed in more detail in the following sections. 
To mention other relatively generic approaches, Flekk{\o}y 
et al. \cite{flekkoy99a,flekkoy00a} have recently introduced 
a hybrid method to combine particle and continuum level 
descriptions. This framework allows a derivation of the DPD 
model and in practice uses a Voronoi tessellation based 
technique for simulations of coarse-grained particles at 
different levels. Another very promising  approach was 
presented by Malevanets and Kapral 
\cite{malevanets99a,malevanets00a} who proposed a framework 
to couple a molecular level description with a mesoscale 
treatment of solvent that conserves hydrodynamics. This 
approach seems particularly appealing for studying dilute 
systems with hydrodynamics.

A lot of impetus for multiscale modeling has come from 
polymer research due to both fundamental and practical 
reasons. Akkermans and Briels \cite{akkermans00a} applied
a projector operator formalism based approach to coarse-grain 
polymer chains using a microscopic model as a starting point. 
Their approach is appealing and deserves further studies 
since it accounts for the fluctuating forces at the 
coarse-grained level. Bolhuis et al. \cite{louis00a,bolhuis01a} 
used the Ornstein-Zernike equation \cite{Hansen86} with a 
hypernetted-chain closure to invert the pair correlation 
function in order to derive effective pair potentials for 
neutral polymers in a solution. The coarse-grained polymers 
were represented as soft colloidal particles. Their approach 
is based on a theorem proved by Henderson in 1974 \cite{hend74} 
that stipulates a unique, point-to-point correspondence between 
pair potentials and radial distribution functions $g(r)$. The 
importance of this theorem becomes clear when one notices that 
for any $g(r)$ under given conditions there is a unique pair 
potential that includes corrections from the many-body interactions, 
and that by definition the pair potential then yields the 
original $g(r)$. The problem is thus reduced to inverting 
the $g(r)$ obtained from a microscopic model -- how to obtain 
a coarse-grained model from that information lies at the heart 
of the problem of coarse-graining. At the representational 
level the soft ellipsoid model for polymer melts by Murat 
and Kremer \cite{murat98a} is very close to the one of Bolhuis 
et al. \cite{louis00a,bolhuis01a}, but instead of using the 
pair correlation function Murat and Kremer base their approach 
on separating the free energy into intra- and interchain parts. 
In an attempt to model more realistic systems, Reith et al. 
\cite{reith00a} introduced a coarse-graining procedure that 
starts from a detailed chemical description of a polymer. In 
order to obtain a coarse-grained approach, a simplex algorithm 
based optimization is performed to obtain a coarse-grained 
force field. The structural features produced by this approach 
are in good agreement  with experimentally observed ones. For 
a more complete overview of the current coarse-graining 
approaches, see e.g. Ref.~\cite{mplathe00} and references 
therein.

Our approach is also based on inverting the radial distribution 
function but instead of the integral equation based approach of 
Bolhuis et al. \cite{louis00a,bolhuis01a} we use the inverse 
Monte Carlo procedure of Lyubartsev and Laaksonen \cite{LL95} 
which will be discussed in the following section. This 
{\it systematic} coarse-graining method is then combined with 
a {\it momentum conserving} dissipative particle dynamics (DPD) 
thermostat. Our approach consists of three conceptually simple 
steps. First, we use a detailed atomistic description and perform 
MD simulations. From these simulations we compute radial 
distribution functions $g(r)$ between different atoms, molecules, 
or molecular groups. Second, we apply the Inverse Monte Carlo 
\cite{LL95} procedure to invert the radial distribution functions. 
This yields effective interaction potentials $V^{\rm eff}(r)$ 
between the selected interaction sites. Finally, we employ these 
effective, coarse-grained potentials in the DPD algorithm to 
study the large-scale properties of systems in question. Note 
that since DPD satisfies momentum conservation, the present 
approach allows studies of a given system with full 
hydrodynamics.

Preliminary results of this work were recently reported 
in Ref.~\cite{karttunen02}, in which we presented only the 
most essential ideas of this approach. Our aim in this article 
is to provide one with a comprehensive discussion of issues 
related to coarse-graining and mesoscopic simulations in 
terms of the DPD method. In particular, we discuss in detail 
how atomistic MD simulations can be coupled to DPD by the 
Inverse Monte Carlo method. To this end, we use NaCl as our 
model system. Even in this simple system, and with a very 
modest degree of coarse-graining, we obtain a computational 
speed-up of one order of magnitude. Our procedure is 
well-defined and truly allows easy and controllable tuning 
of the level of coarse-graining. Finally, we show through 
coarse-grained DPD simulations that this approach is physically 
sound and provides results in excellent agreement with the MD 
simulations and experiments.

This paper is organized as follows. 
First, in Sect.~\ref{sec:methods}, we describe the 
methods: The Inverse Monte Carlo procedure and the DPD algorithm. 
In Sect.~\ref{sec:results} we present the results and finally, 
in Sect.~\ref{sec:concl}, we discuss the results and some general 
aspects related to the general applicability of the method.

\section{Methods and models} 
\label{sec:methods}

As discussed in the Introduction, detailed atomistic simulation 
techniques are suitable for studies of microscopic features 
of soft matter systems, e.g. the excited 
states of a chromophore where the time scales of interest are 
within a few hundreds of femtoseconds. For studies of systems 
that are characterized by large time and length scales, such as 
a protein in water solution over tens or hundreds of nanoseconds, 
however, atomistic simulation techniques are not very feasible. 
Our aim in this section is to discuss ways how this problem 
can be resolved by coarse-graining microscopic systems. In 
particular, we concentrate on an approach based on solving the 
``inverse problem'' which yields effective interaction potentials 
which in turn can be coupled to mesoscopic simulation techniques such 
as the dissipative particle dynamics method discussed below. 
Full details of the methods can be found in Refs.~\cite{LL95} 
and \cite{Bes00,Vattulainen02}, respectively.

\subsection{The Inverse Problem in statistical mechanics} 
 
In the standard statistical mechanical description of soft 
matter systems one begins from a formulation of a model, 
which is usually given in terms of a Hamiltonian or interaction 
potentials between atoms and molecules. When a Hamiltonian 
is specified, one can use different numerical methods to 
compute canonical averages such as energies, distribution 
functions, and response functions which can be compared 
with experiments. As one has an explicit expression for the 
canonical averages, their calculation is in principle a 
fairly straightforward task, although in many cases it 
may be computationally a very demanding one.

Occasionally one is interested in the solution of the inverse 
problem, i.e., how to deduce information about molecular 
interactions from the canonical averages which may be known from
experiments. More specifically, the question is how to 
reconstruct the Hamiltonian based on results for the canonical 
averages. Many experimental properties can be chosen as 
a starting point to this end. Among the most important ones 
for the liquid state are the radial distribution functions $g(r)$, 
which may be known from the structure factor measurements using 
X-ray or neutron scattering \cite{Hansen86}. In 1974, Henderson 
proved a theorem \cite{hend74} stipulating a unique, point-to-point 
correspondence between pair potentials and radial distribution 
functions. In essence it states that for a given system under 
given conditions of temperature and density, two pair potentials 
which give rise to the same radial distribution function cannot 
differ more than by an additive constant. As this constant can be 
defined from the condition that the interaction potential tends 
to zero at an infinite distance, the potential is uniquely 
defined. In practice the solution of an inverse problem is not 
a straightforward calculation, however, since $g(r)$ does not 
provide one with a formal expression for the potential. Therefore, 
some special techniques are required for the solution of the 
inverse problem.

In 1988, McGreevy and Pusztai suggested a Reverse Monte Carlo 
(RMC) method in which the starting point for a simulation was the 
radial distribution function \cite{RMC88}. In this method, Monte 
Carlo simulations are carried out without any prior knowledge of 
the interaction potential to fit $g(r)$  that serves as an input. 
The set of configurations obtained in RMC may be used for further 
structural analysis, for example for calculation of three-dimensional 
spatial or orientational correlations. However, as the interaction 
potential is not recovered, the inverse problem is not solved 
completely, and it is not possible to calculate thermodynamical or 
dynamical properties using this approach. For the same reason 
this approach in not suitable for the development of 
coarse-grained models.

In another approach, Reatto et al. \cite{reatto86} used the 
Hypernetted Chain (HNC) approximation to solve the inverse 
problem. This and similar approaches, based on some closure 
of the Ornstein-Zernike equation, have been used in a number 
of works during the last decade to compute interaction potentials 
from radial distribution functions \cite{rosenfeld97,gonzalez98}. 
It should be noted, however, that computations within the HNC 
theory are feasible only for relatively simple models. Moreover, 
although yielding sometimes very accurate results \cite{LyMar02}, 
the HNC theory is not an exact mathematical solution of the  
statistical-mechanical problem, and its accuracy should be 
investigated carefully in every specific case. Other works 
devoted to the inverse problem are listed in 
Refs.~\cite{ostheimer89,soper96,toth00}.

A practical way to solve the inverse problem has been 
suggested by Lyubartsev and Laaksonen \cite{LL95}. This is the 
method we will concentrate on in the following, and we will 
use it to compute effective potentials from radial distribution 
functions obtained from detailed MD simulations. The effective 
potentials then allow us to to study the large-scale properties 
of a given model system through coarse-grained DPD simulations.

\subsection{The Inverse Monte Carlo Method -- A Simple Case} 
 
To illustrate the Inverse Monte Carlo method, we consider 
the case of a single-component system consisting of identical 
particles with pairwise interactions. The general case of 
a multicomponent system is a straightforward extension of it. 
The Hamiltonian for this system is given as 
\begin{equation} 
H =  \sum_{i,j}V(r_{ij}),           
\label{ham}  
\end{equation} 
where $V(r_{ij})$ is the pair potential and $r_{ij}$ is 
the distance between the interaction sites $i$ and $j$. 
Let us assume that we know the radial distribution 
function $g(r_{ij})$. Our aim is now to find the 
corresponding interaction potential $V(r_{ij})$.

Let us introduce the following grid approximation to 
the Hamiltonian,
\begin{equation} 
\widetilde{V}(r) = V(r_{\alpha}) \equiv V_{\alpha} 
\end{equation} 
for 
\begin{equation} 
r_{\alpha} - \frac{1}{2M} < r < r_{\alpha} + \frac{1}{2M} 
\mbox{\hspace{0.5cm} and  \hspace{0.5cm}} 
r_{\alpha} = ({\alpha} - 0.5) \, r_{cut} / M ,  
\end{equation} 
where $ \alpha = 1,\ldots, M $, and $M$ is the number of grid 
points within the interval $[0,r_{cut}]$, and $r_{cut}$ 
is a chosen cut-off distance. Then, the Hamiltonian 
in Eq.~(\ref{ham}) can be rewritten as     
\begin{equation}  
H = \sum_{\alpha} V_{\alpha} S_{\alpha},               
\label{ham_gr}  
\end{equation}  
where $S_{\alpha}$ is the number of pairs with interparticle 
distances inside the $\alpha$-slice. Evidently, $S_{\alpha}$ 
is an estimator of the radial distribution function:   
$\langle S_{\alpha}\rangle = 4{\pi}r^2g(r_{\alpha})N^2/(2V)$.   
The average values of $S_{\alpha}$ are some functions of the potential  
$V_{\alpha}$ and can be written as an expansion 
\begin{equation}  
\Delta\langle S_{\alpha}\rangle 
  = \sum_{\gamma}\frac{\partial \langle  
     S_{\alpha}\rangle }{\partial V_{\gamma}} \Delta V_{\gamma} 
        + \mathcal{O}(\Delta V^2).  
\label{dif}  
\end{equation}  
The derivatives   
$\partial \langle S_{\alpha}\rangle / \partial V_{\gamma}$   
can be expressed using exact statistical mechanics 
relationships \cite{LL95}
\begin{equation}  
\begin{array}{lcl} 
{\displaystyle 
\frac{\partial \langle S_{\alpha}\rangle }{\partial V_{\gamma}} } & = & 
{\displaystyle 
\frac{\partial}{\partial V_{\gamma}} \frac 
{\int {\rm dq} \, S_{\alpha}(q) 
      \exp (-\beta \sum_{\lambda}K_{\lambda}S_{\lambda}(q))} 
{\int {\rm dq} \, \exp (-\beta \sum_{\lambda}K_{\lambda}S_{\lambda}(q))} 
}~\vspace{0.2cm} \\
& = & 
{\displaystyle 
 - \frac{ \langle S_{\alpha}S_{\gamma}\rangle -   
\langle S_{\alpha}\rangle \langle S_{\gamma}\rangle } {k_B T}  . } 
\end{array} 
\label{matr}    
\end{equation}  
Equations~(\ref{dif}) and (\ref{matr}) allow us to find 
the interaction potential $V_{\alpha}$ iteratively from the 
radial distribution functions  $\langle S_{\alpha}\rangle$. 
Let $V_{\alpha}^{(0)}$ be a trial potential for which 
a natural choice is the potential of mean force (PMF) 
\begin{equation}  
V_{\alpha}^{(0)} = - k_B T \, {\ln} \, g(r_{\alpha}),
\end{equation} 
By carrying out standard Monte Carlo simulations, one 
can evaluate the averages $\langle S_{\alpha}\rangle$ 
and their deviations from the reference values 
$S_{\alpha}^{\ast}$ defined from the given radial 
distribution function as $\Delta \langle S_{\alpha}\rangle ^{(0)} =   
\langle S_{\alpha}\rangle ^{(0)} - S_{\alpha}^{\ast}$. 
By solving the system of linear equations [Eq.~(\ref{dif})] 
with coefficients defined by Eq.~(\ref{matr}), and omitting 
terms $\mathcal{O}({\Delta}V^2)$, we obtain corrections to 
the potential ${\Delta}V_{\alpha}^{(0)}$. The procedure is 
then repeated with the new potential 
$V_{\alpha}^{(1)} = V_{\alpha}^{(0)} + {\Delta}V_{\alpha}^{(0)}$
until convergence is reached. The whole procedure resembles 
a solution of a multidimensional non-linear equation using 
the Newton-Raphson method.

If the initial approximation of the potential is poor, 
some regularization of the iteration procedure is needed. 
In this case we multiply the required change of the radial 
distribution function by a small factor that is typically 
between 0 and 1. By doing so, the term $\mathcal{O}({\Delta}V^2)$ 
in Eq.~(\ref{dif}) can be made small enough to guarantee 
convergence of the whole procedure, although the number 
of iterations will increase.

The above algorithm has also the advantage that it provides
us with a method to 
evaluate the uncertainty of the inverse procedure. The 
radial distribution function has normally some uncertainty. 
An analysis of the eigenvalues and eigenvectors of the matrix 
in Eq.~(\ref{matr}) allows one to make conclusions of which 
changes in $g(r)$ correspond to which changes in the potential. 
For example, eigenvectors with eigenvalues close to zero 
correspond to changes in the potential which have almost 
negligible effect on the radial distribution function. The 
presence of these small eigenvalues makes the inverse problem 
not well-defined, however. In some cases, such as for liquid 
water, that may pose serious problems in the inversion 
procedure \cite{LL_CPL00}.

It is instructive to note a relation between the present 
approach and the renormalization group Monte Carlo method 
\cite{swendsen79,pawley84} used to study phase transitions 
in lattice models (e.g., polymer models and Ising models 
for ferromagnets) as well as in the quantum field theory.  
The renormalization procedure introduces a scale change in 
the system.  During this procedure one consecutively obtains 
a more and more coarse-grained description of the system. 
Equations~(\ref{dif}) and (\ref{matr}) were in fact used by 
Swendsen et al. \cite{swendsen79,pawley84} to describe how 
the parameters of a Hamiltonian change during the renormalization 
procedure. It now appears that the applications of this method 
are more general than the original lattice systems near a phase 
transition, and cover even soft matter problems allowing us 
to ``renormalize'' Hamiltonians of molecular systems in such 
a way that only the degrees of freedom of primary interest 
are kept in the coarse-grained system.

\subsection{Dissipative Particle Dynamics (DPD)}

Dissipative particle dynamics was introduced in 1992 by
Hoogerbrugge and Koelman for simulations of hydrodynamic 
phenomena in complex fluids \cite{hoogerbrugge92a,koelman93a}. 
The original formulation did not obey detailed balance, 
though, and in 1995 Espa{\~n}ol and Warren \cite{espanol95a} 
formulated a new DPD algorithm which they showed to be fully 
consistent with statistical mechanics. This algorithm is 
nowadays known as DPD. In the following we present the basic 
DPD formalism and discuss some of its features that are 
relevant to the present work. 
\begin{figure*}[bt!]
\includegraphics[width=15cm]{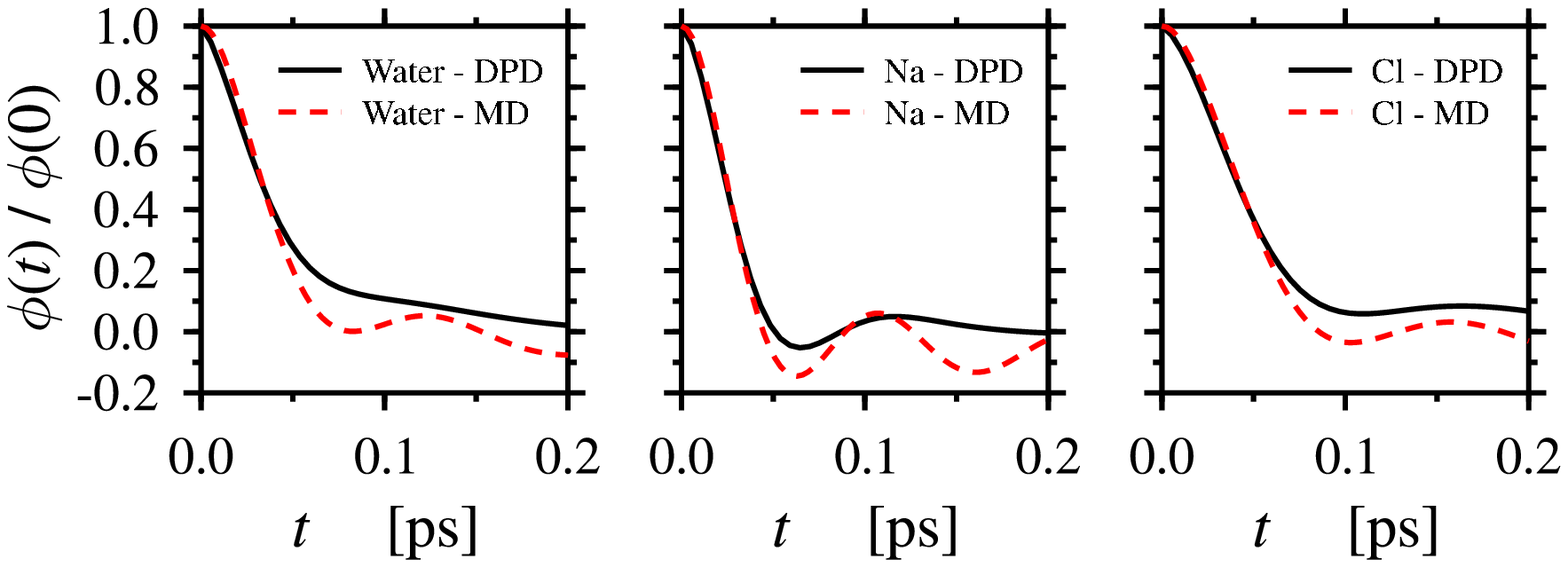}
\caption{The decay of the single-particle velocity autocorrelation 
         function $\phi(t)$ at early times. Results shown here are 
         for water, Na$^{+}$ and Cl$^{-}$ ions. As the data 
         illustrates, the early-time decay of $\phi(t)$ is 
         essentially identical between MD and DPD simulations for 
         $\gamma = 0.72$ used in the DPD simulations.}
\label{figure_vacf} 
\end{figure*}

In the basic formulation of DPD, the interactions are pairwise 
additive and the force exerted by particle $j$ on particle $i$ is 
given as a sum of conservative, dissipative, and random forces through 
$\vec{F}_{ij} = \vec{F}_{ij}^C + \vec{F}_{ij}^D + \vec{F}_{ij}^R$, 
respectively. These forces are typically given as 
  \begin{eqnarray}
  \label{DPD_forces}
  \vec{F}_{ij}^C & = & F_{ij}^{(c)} (r_{ij}) \,  
   \vec{e}_{ij} , \label{forces:1} \\
  \vec{F}_{ij}^D & = & - \gamma \, \omega^{D} (r_{ij}) 
   (\vec{v}_{ij} \cdot  \vec{e}_{ij})\, \vec{e}_{ij} , \label{forces:D} \\
  \vec{F}_{ij}^R & = & \sigma \, \omega^{R} (r_{ij}) \, \xi_{ij} \, 
   \vec{e}_{ij} , \label{forces:2}
  \end{eqnarray}
where $\vec{r}_{ij} = \vec{r}_i - \vec{r}_j$ is the position 
vector connecting two particles, $r_{ij} = | \vec{r }_{ij} |$ 
is the interparticle distance, $\vec{e}_{ij} = \vec{r}_{ij} / r_{ij}$ 
is the corresponding unit vector, and  
$\vec{v}_{ij} = \vec{v}_i - \vec{v}_j$ is the relative 
velocity of particles $i$ and $j$. The terms $\xi_{ij}$ are 
symmetric random variables with zero mean and unit variance, 
and are independent for different pairs of particles and 
different times. The conservative force is essentially given 
by $F_{ij}^{(c)} (r_{ij})$. Constants $\gamma$ and $\sigma$ 
give the amplitudes of the dissipative and random forces, 
respectively, and $\omega^{D}$ and $\omega^{R}$ are the 
corresponding weight functions.

Let us now return to the different forces. 
The DPD formulation does not specify the form of the 
conservative force. The most common choice in standard DPD 
simulations is $F_{ij}^{(c)} (r_{ij}) = 
\alpha_{ij} \, ( 1 - r_{ij} ) / r_c$ for $r \le r_c$ and 
$F_{ij}^{(c)} (r_{ij}) = 0$ otherwise. In other words, the 
conservative force is derived from a soft pair potential 
$U = \alpha_{ij} ( 1 - r_{ij} / r_c)^2 / 2$, where $r_c$ is 
the cutoff distance. Interactions between different types 
of coarse-grained particles can then be described by varying 
the amplitude of the conservative interaction 
$\alpha_{ij}$ \cite{groot97a}.

The above is the preferred formulation for the conservative 
force when generic, simple, and soft interactions are adequate. 
It was shown by Forrest and Suter in 1995 that coarse-graining 
of molecular representation leads to this type of softening 
\cite{For95}. However, in many cases this formulation is of too 
generic nature. Although it is possible to use mean field 
theories such as the Flory-Huggins theory for polymers to 
find the amplitudes for the conservative forces between 
different types of particles \cite{groot97a}, there are cases 
where this approach does not retain enough information about 
the actual character of the different atoms, molecules, or 
interactions (see discussion below).

The weight functions for the dissipative and random forces, 
$\omega^{D}$ and $\omega^{R}$, cannot be chosen arbitrarily. 
This is easy to understand by the intuitive argument that 
the thermal heat generated by the random force must be 
balanced by dissipation. Espa{\~n}ol and Warren \cite{espanol95a}
studied this situation analytically and derived a
fluctuation-dissipation relation connecting the weight 
functions and amplitudes of the dissipative and random 
forces via 
  \begin{equation} 
  \label{eq:fluc1}
  \omega^{D} (r_{ij})  =  [ \omega^{R} (r_{ij}) ]^2 
  \mbox{\hspace{0.5cm} and \hspace{0.5cm}} 
  \sigma^2  =  2 \, \gamma \, k_B T. %%%^{\ast}
  \end{equation}
These conditions guarantee convergence to the canonical 
ensemble as required. It is important to notice that 
$\omega^{D}$ and $\omega^{R}$ are completely decoupled from 
the conservative part. In other words, we can consider the 
DPD algorithm as a momentum conserving thermostat for an 
arbitrary conservative potential. This is in fact the way 
how we apply the DPD algorithm in our coarse-grained simulations: 
The conservative potential is defined through the Inverse Monte 
Carlo procedure and then used in the DPD algorithm.

Finally, to study the dynamics of the system one needs to 
evolve the system in time. In DPD this is simply done by 
integrating the Newton's equations of motion. In our 
simulations we used a velocity-Verlet based algorithm adapted 
for DPD (sometimes called DPD--VV) \cite{Bes00,Vattulainen02}. 

\subsection{Combining MD and DPD}

With the above definitions we now have all the tools for 
coarse-graining. We remove fast internal and orientational 
degrees of freedom from all water molecules, and represent 
water molecules as one-site particles interacting with other 
water particles and ions by spherically symmetric potentials. 
In comparison to the original all-atom model, we have 
approximately one third of the degrees of freedom left. This 
can be interpreted as intermediate level of coarse-graining.

Without any loss of generality we can choose the weight 
functions to be of the standard form 
$\omega^{R}(r_{ij}) = 1 - r_{ij} / r_c$ as discussed above. 
This choice has been made invariably in 
DPD simulations. Since the Inverse Monte Carlo 
procedure provides us with the effective potentials, 
thus yielding us $ \vec{F}_{ij}^C(r) $, the 
only task left is to ensure that the MD and DPD systems 
correspond to each other. In this study, that was done by adjusting the 
amplitude of the dissipative force $\gamma$ which determines 
the dissipation rate in the DPD system. Here, $\gamma$ was 
determined by requiring the short-time decay of the 
single-particle velocity autocorrelation function 
\begin{equation} 
\phi(t) \equiv \langle \vec{v}_i(t) \cdot \vec{v}_i(0) \rangle 
\end{equation} 
to be approximately the same in both MD and DPD systems 
as shown in Fig.~\ref{figure_vacf}. It shows how the 
early-time decay of $\phi(t)$ can be matched, leading 
to a value of $\gamma = 0.72$ for the DPD model. The 
long-time behavior of the velocity autocorrelation function 
between MD and DPD is different, however. This is an obvious 
result since some microscopic degrees of freedom have been 
coarse grained and thus effects due to hydrogen bonds, for 
example, are {\it implicitly} included in the effective 
interactions. This leads to an enhanced diffusion rate at 
intermediate times, and is demonstrated in Fig.~\ref{figure_vacf} 
as a positive tail for the DPD particles.

This approach is meaningful since ({\it i\,}) it makes sure 
that the early-time dynamics is described properly, while 
({\it ii\,}) it does not fix the tracer diffusion coefficient. 
One should notice that the tracer diffusion coefficient is an 
integral over the velocity correlation function over long times, 
and in hydrodynamic systems the long-time tail is important. 
With this choice, we can assume the diffusion coefficient to be 
an independent quantity so that its behavior, found by DPD, can 
be compared with both MD simulations and experiments. Fixing 
$\gamma$ thus determines the transport properties of the 
system.

\section{Results}
\label{sec:results}

\subsection{Constructing the potentials}

One of the typical simplifications used in 
molecular simulations is the replacement of solvent molecules 
by continuum media. For example, in the primitive electrolyte 
model, ions in water are substituted by charged spheres 
moving in dielectric media with the dielectric constant 
set to about 80. This is a serious simplification at small 
distances (a few {\aa}ngstr\"oms) where it is impossible to 
define a dielectric constant. Besides this, in the primitive 
electrolyte model the ion radius is an adjustable parameter 
without any clear physical meaning. A better model for 
effective ion-ion interactions in an aqueous solution must 
take into account the solvation structure around the ions. 
Practically, effective solvent-mediated ion-ion potentials 
may be constructed by the Inverse Monte Carlo method from 
ion-ion radial distribution functions generated in high-quality 
all-atomic molecular dynamics simulations \cite{LL95,LyMar02,LL97}. 
This is the approach we used in obtaining the effective 
potentials.

\begin{figure}%[bt!]
\includegraphics[width=8cm]{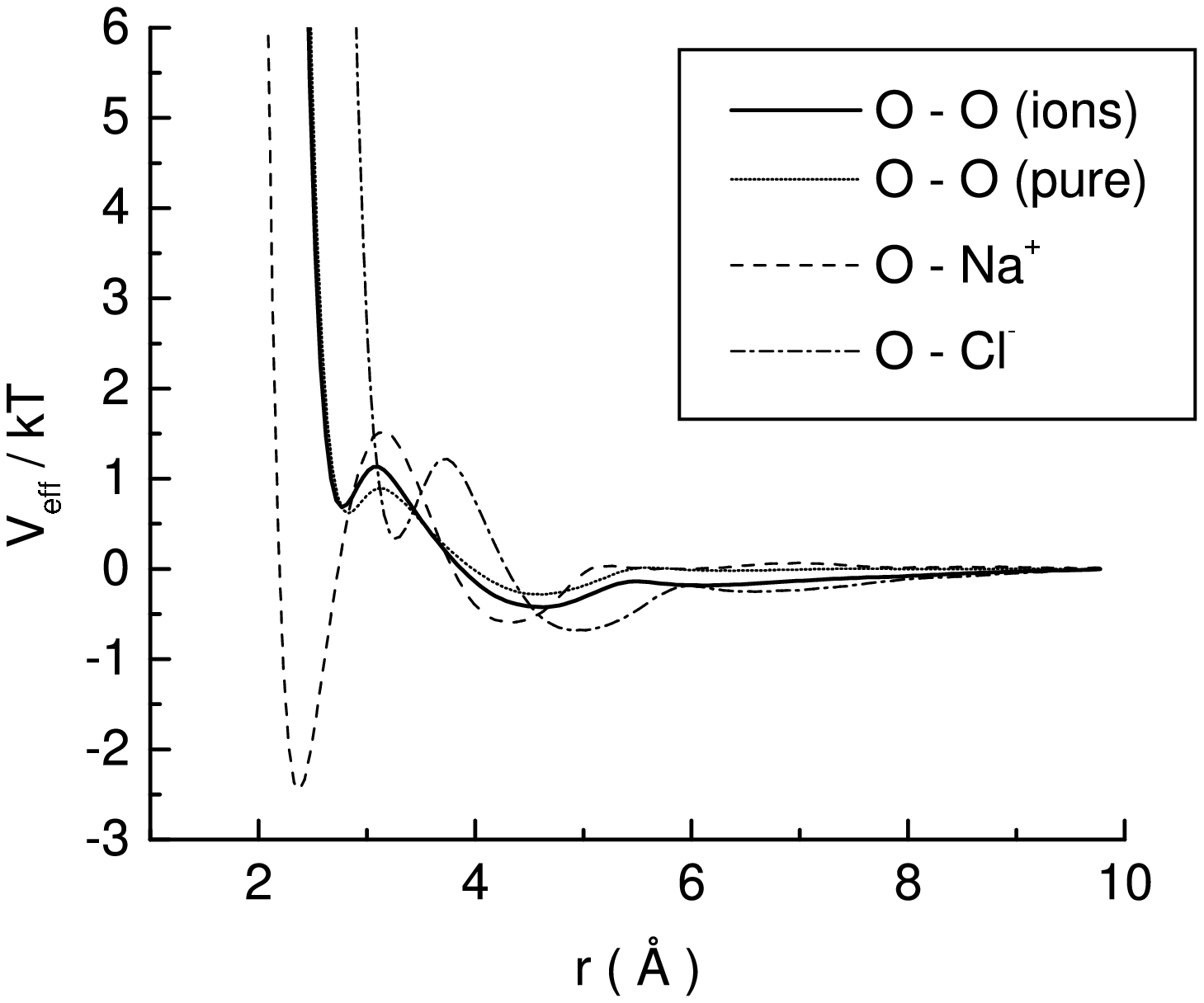}
\caption{Effective potentials for coarse-grained water in 
         NaCl solution. Thin dotted line is water-water 
         effective potential in pure water. ``O'' refers 
         to an oxygen atom in the water molecule.}
\label{O_pot}
\end{figure}

\begin{figure}%%[bt!]
\includegraphics[width=8cm]{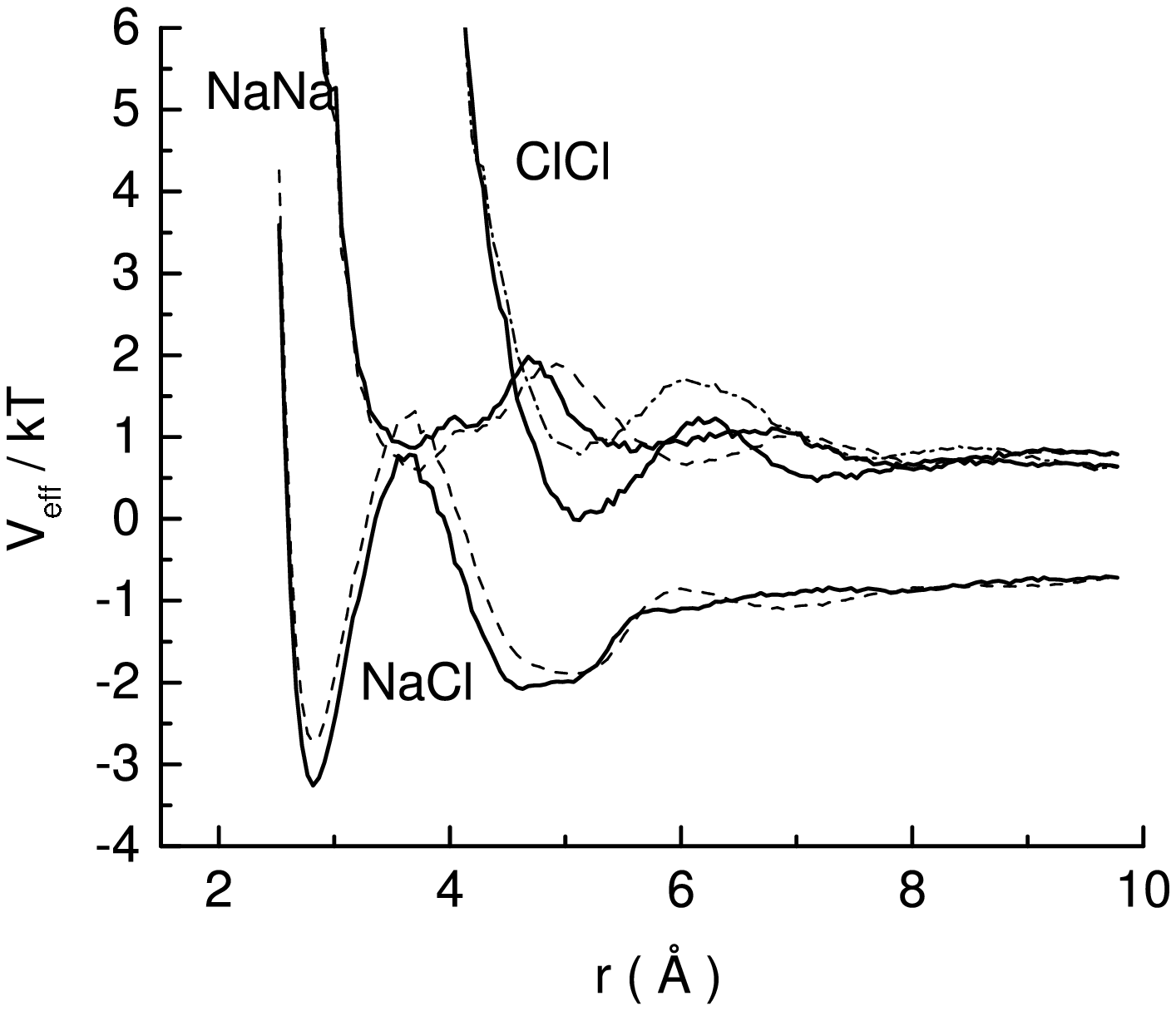}
\caption{Ion-ion effective potentials in aqueous solution. 
         Solid lines describe  potentials in presence of 
         one-site water molecules (this work), while dashed 
         lines are for effective potential without water from 
         Ref.~\cite{LL97}.}
\label{ion_pot}
\end{figure}

The MD simulations were performed in the NVT ensemble 
using the flexible SPC water model \cite{tourah} and 
Smith--Dang parameters for Na$^+$ and Cl$^-$ ions \cite{SD94}, 
i.e., the Lennard-Jones parameters for the sodium ions    
were $\sigma = 2.35$\,{\AA} and $\varepsilon = 0.544$\,kJ/mol, 
and for chloride $\sigma = 4.4$\,{\AA} and $\varepsilon = 0.419$\,kJ/mol.
In the results reported here we set the temperature to 300~K.
The electrostatic interactions were taken into account by Ewald 
summation. Other simulation details are described in 
Refs.~\cite{LL95,LL97}. From the MD simulations, the radial 
distribution functions between different pairs of particles 
were calculated and fed as an input in the Inverse Monte Carlo 
procedure. The results are shown in Figs.~\ref{O_pot} 
and \ref{ion_pot}.

Figure~\ref{O_pot} shows effective potentials between water 
molecules, presented as one-center particles, and between 
other water molecules and ions. For comparison, water-water 
effective potential calculated from MD simulations for 
{\em pure} water is also displayed. It is interesting to 
note that the presence of ions has practically no effect 
on the water-water effective potential.

Figure~\ref{ion_pot} displays the effective potentials between 
the ions. They are compared to the effective potentials 
calculated without water \cite{LL95,LL97}. The effective 
potentials are very similar in the two cases both of them 
having an oscillating character with one or two oscillations. 
The potential approaches the primitive model potential with 
dielectric constant close to 80. At distances above 10~{\AA} 
the effective potential coincides almost perfectly with the 
Coulombic potential.

Within the inverse Monte Carlo procedure, calculation of 
effective potentials without water is much easier than in 
the presence of water. On the other hand, the presence of 
particles representing water is necessary in DPD simulations. 
Our test studies have shown that while the use of effective 
potentials calculated without water may give qualitatively 
satisfactory results, inclusion of water in the Inverse Monte 
Carlo simulations essentially leads to an improvement of 
results, especially in the case of dynamical properties.

It is worth pointing out that the effective 
potential differs from the potential of mean force 
defined as \,$\psi _{PMF} = -k_B T \, \ln\,g(r)$, which 
corresponds to the Kirkwood approximation for the $n$-particle 
correlation function. The potential of mean force is screened by 
the other ions in a system and decays as $(1/r)e^{(-r/r_D)}$, i.e., 
not as the $1/r$ Coulombic potential. The potential of mean force 
is therefore not a very suitable choice to present ion-ion 
interactions within a continuum solvent model at finite ion 
concentrations even in a coarse-grained solvent system. The 
effective potentials include a contribution from Coulombic 
forces, however. 

\subsection{Coarse-grained simulations}

In our first paper using this approach \cite{karttunen02}, we 
showed that the radial distribution functions computed from 
the coarse-grained DPD simulations for the pairs between the 
Na$^{+}$ and Cl$^{-}$ ions agreed with those obtained from the 
all-atom MD simulations. In addition to the static properties, 
the diffusion coefficients were shown to be in good agreement 
between the MD and DPD results at various NaCl concentrations 
between 0.5 M and 4.1 M. What makes those results very 
interesting is the fact that, for all the DPD simulations, 
we used the effective potentials obtained at one single 
NaCl concentration only. In that case, the effective potentials 
were obtained from MD simulations at a salt concentration of 
0.87 M and then used in DPD simulations for various salt 
concentrations between 0.5 M and 4.1 M. When compared with 
MD simulations, the agreement was found to be excellent. 
Here, we present results for static correlations,
coordination numbers, velocity autocorrelation functions,
and viscosity.  

\begin{center}
\begin{table}[htb]
\begin{tabular}{|l|l|l|l|l|l|} 
\hline
  & molarity:   & 0.55 M   & 0.87 M   & 2.2 M   & 4.1 M \\ \hline \hline
MD: & Na$^+$:     & 5.8      & 5.5      & 5.42    & 5.3 \\ \hline
  & Cl$^-$:     & 6.9      & 6.8      & 6.9     & 7.1 \\ \hline\hline
DPD: & Na$^+$:    & 5.67    & 5.66      & 5.55    & 5.40 \\\hline
  & Cl$^-$:     & 6.91     & 6.96     & 6.97    & 7.00 \\ \hline
\end{tabular}
\caption{Coordination numbers obtained from MD and DPD simulations 
         at various salt concentrations. The error bars for both 
         cases are $\pm 0.05$. The MD data is from Ref.~\cite{sasha96}.}
\label{tab:coord}
\end{table}
\end{center}

\begin{figure*}%%[htb]
\includegraphics[width=13.5cm]{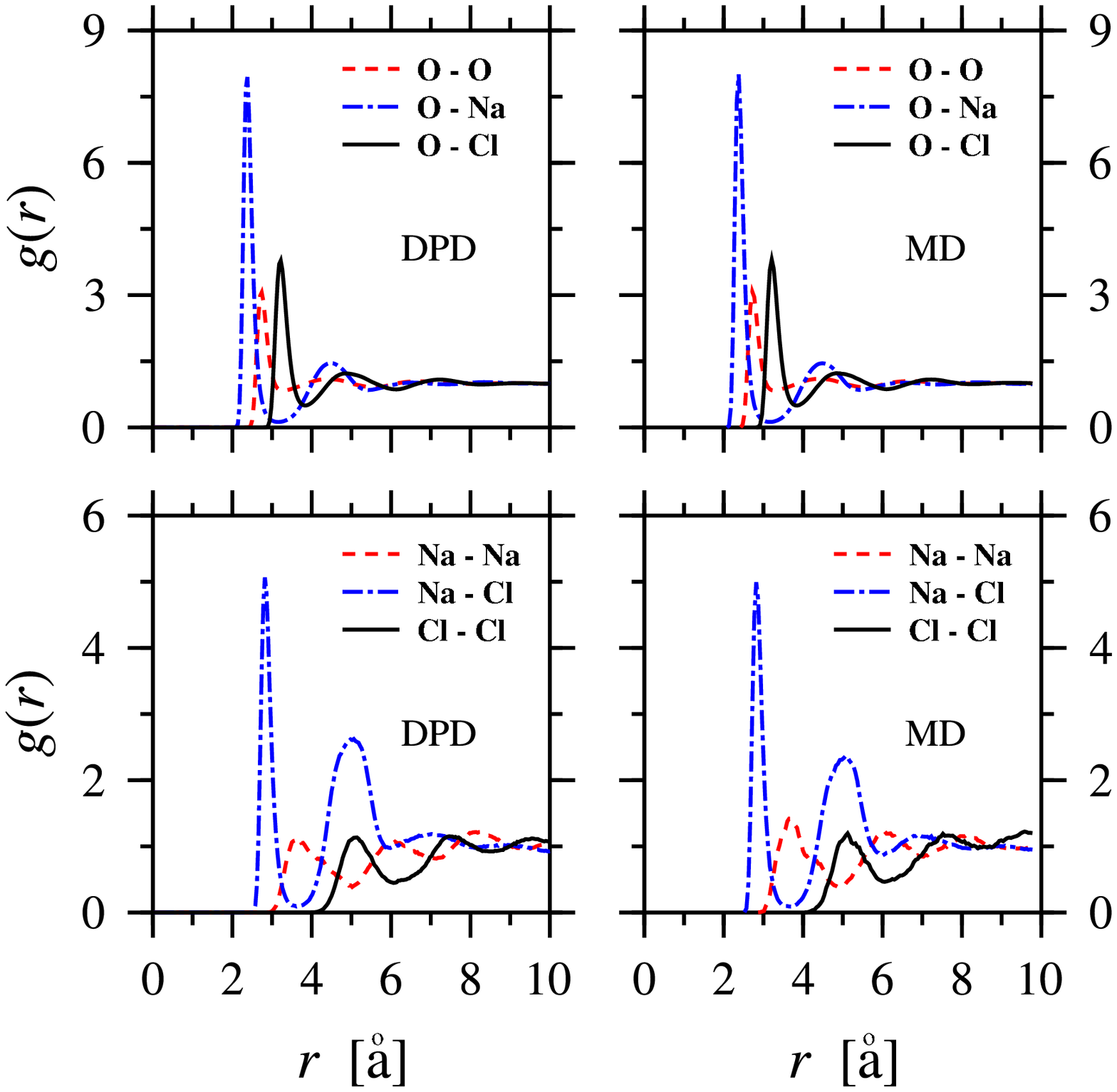}
\caption{Radial distribution functions $g(r)$ for different 
         pairs of particles in the aqueous NaCl system studied 
         by both DPD (shown on the left) and MD (on the right). 
         The studies were made at a salt concentration of 0.87 M. 
         ``O'' refers to the ogygen atom in the water molecule. 
         Note the agreement between MD and DPD results.}
\label{figure_grs}
\end{figure*}

It is important to notice that due to the self-consistency 
conditions for the effective potentials, the DPD simulations 
will always produce the correct pair correlation behavior at 
the point in phase space where the effective interactions 
were determined in the first place. As an example, this is 
demonstrated in Fig.~\ref{figure_grs} for all different kinds 
of pairs of particles studied in this work. It is very relevant 
to ask, however, how the effective interactions change when 
system variables such as the salt concentration or the 
temperature of the system change. We have explored this 
matter through combined MD and DPD simulations and found that 
for a wide range of salt concentrations in the present system 
it is {\it not} necessary to recalculate the effective potentials. 
This is demonstrated by the data in Table~\ref{tab:coord} which 
compares the coordination numbers obtained from DPD and MD 
simulations for $g(r)$. In the DPD simulations we have again 
used the effective potentials obtained at 0.87 M. We find 
that both MD and DPD follow the same qualitative behavior. 
For sodium the coordination number decreases with increasing 
salt concentration whereas for chloride we find an opposite trend. 
Furthermore, the quantitative agreement between the MD and DPD 
results is remarkably good. 

\begin{figure}[htb]
\includegraphics[width=12cm]{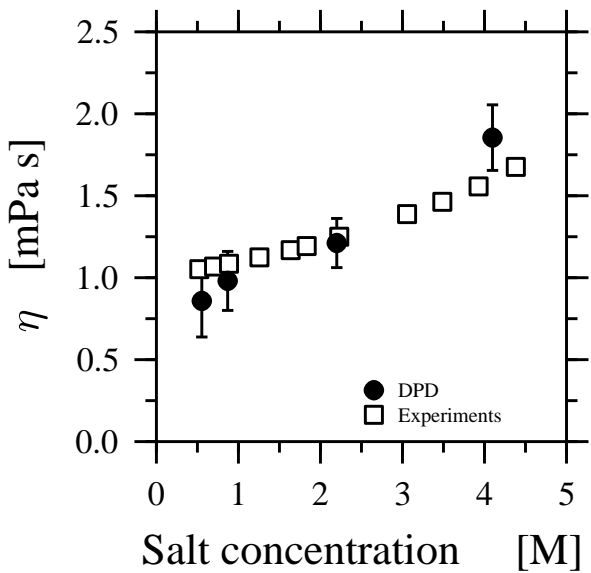}
\vspace*{-1.5cm}
\caption{The shear viscosity coefficient $\eta$ from coarse-grained DPD 
         simulations and experiments. The experimental data is from 
         Ref.~\cite{crc}. The data by DPD has been scaled 
         to allow comparison.}
\label{fig:shear}
\end{figure}

Our finding that the effective interactions depend only weakly 
on thermodynamic conditions is also supported by the present 
results for shear viscosity (see below) and earlier studies 
of the same system for tracer diffusion \cite{karttunen02}. 
In Ref.~\cite{karttunen02}, we found that DPD simulations with 
effective potentials yielded tracer diffusion coefficients in good 
agreement with MD simulations. Further support for this result is 
given by previous work \cite{LyMar02,LL97}, where the dependence 
of effective potentials on the salt concentration was studied in 
detail. It was found that effective potentials depend 
very little on the salt concentration: An increase in salt 
concentration leads to a slow decrease in the dielectric 
permittivity.

Our results thus suggest that the effective interactions 
change only slightly when the system parameters are varied, 
as is here the case for salt concentration. However, while this 
result holds true in the present system for an aqueous solution 
of monovalent ions, the generality of this finding remains an 
open question. Further studies of divalent systems and more 
complex molecules, among others, are therefore called for.

Next, we determined the shear viscosity using a Green-Kubo relation. 
As with the other quantities, the shear viscosity should be 
compared to results from the MD simulations. However, since the 
shear viscosity coefficient is a collective quantity, meaning 
that all particles in a system give rise to a single sample, an 
accurate determination of the shear viscosity coefficient through 
MD simulations is both difficult and very time-consuming. Thus 
it was not done in the present case, and as a matter of fact, 
to the best of our knowledge, there are no published reports of 
MD simulations of shear viscosity in NaCl solutions. Thus, we 
compare our DPD data to experimental results \cite{crc} instead. 
The results shown in Fig.~\ref{fig:shear} indicate that the 
qualitative behavior of the shear viscosity coefficient obtained 
through DPD simulations is in good agreement with the experiments. 
This result is truly remarkable as it shows that both the 
equilibrium (static) and dynamic behavior of the system are 
properly described by the coarse-grained approach. This 
provides us with strong support that the present approach 
of coarse graining an MD system and applying the obtained effective 
interactions in DPD simulations is a promising approach indeed.

We would like to note that we have very recently performed 
simulations using LiCl and CaCl$_2$ \cite{terama02} to study 
further aspects of this issue, such as the effects of salt 
concentration, valence, and temperature. Despite the very 
different natures of these solutions (as compared to NaCl), 
the conclusions with regard to the coarse-graining method and 
its general applicability remain the same.

\section{Conclusions}
\label{sec:concl}

In this article we have described in detail a method
that combines the Inverse Monte Carlo method and the 
momentum conserving dissipative particle dynamics 
thermostat for coarse-grained simulations. This approach 
allows systematic coarse-graining of molecular systems
and conserves hydrodynamics. 
Here, the level of coarse-graining has been quite modest 
and the potentials used did not have particularly soft 
cores. This is due to the fact that in this system of water 
and NaCl, we coarse-grained the molecular representation 
of water molecules only. Thus, the corresponding effective 
interactions retained features which are typical for systems 
described on an atomic level. However, it is important to 
notice two facts. First of all, further coarse-graining is 
not possible for this kind of a system if one wishes to 
preserve the identity of individual ions and the essential 
characteristics of water molecules. One could, of course, 
continue the coarse-graining procedure to obtain softer 
potentials by clustering several water molecules together 
and treating the clusters as new ``effective'' solvent particles. 
This would lead to further computational gains but the ions and 
molecules would then lose essential parts of their identity. 
As a result of this the interactions would become softer and
resemble the quadratic potentials used for conservative 
interactions in the original DPD formulation. 
Second, despite the modest level of coarse-graining in this 
work, the computational gain is still remarkable, around 
a factor of 20. This speed-up is due to the fact that no 
explicit Ewald summation was required for the coarse-grained 
system (since the effective potentials include information 
about the electrostatics, and the screening length was 
shorter than the applied cut-off) and that SPC water used 
in molecular dynamics simulations is replaced by a single 
coarse-grained particle with the electrostatic interactions 
included in the effective potentials.

This approach allows one to coarse-grain complex macromolecular 
systems to a  level where a detailed molecular 
description is replaced with a simplified model of coarse-grained 
particles. For example, a lipid-acyl chain  ``$\ldots$\,--\,CH$_2
$\,--\,CH$_2$\,--\,CH$_2$\,--\,CH$_2$\,--\,CH$_2$\,--\,CH$_2$ $\ldots$'' 
can be described as  ``$\ldots$--\,B\,--\,B\,--\,$\ldots$'' 
where each of the beads ``B'' represent 2\,--\,4 methyl groups 
CH$_2$. Since the beads ``B'' describe the interaction site 
for a whole molecular group, it is clear that the interactions 
between different beads will be relatively soft in the spirit 
of the original DPD formulation, yet retaining their characteristic 
features well beyond the mean-field description of the original 
DPD formulation. For macromolecular systems, this coarse-graining 
procedure thus has the potential to lead to a major reduction 
in computational effort as well as to softer interaction potentials. 
The computational gain depends on the type of interactions: 
While here in the case of NaCl we did not need to use Ewald 
summation, the situation may be different in the case of divalent 
ions such as calcium. In addition, one should note that 
coarse-graining of a macromolecular system requires a lot of 
insight, and there is no unique way of selecting the units to
be coarse-grained. Work is in progress to address these issues 
and the coarse-graining of macromolecular systems in general.

The approach presented in this article has close connections 
to other methods as well. The Inverse Monte Carlo procedure has 
the same foundations as the integral equation methods of Bolhuis 
et al. \cite{louis00a,bolhuis01a}. Second, the coarse-graining 
procedure used by Reith, Mayer and M\"uller-Plathe \cite{reith00a} 
has a close resemblance to the Inverse Monte Carlo procedure.
The latter group has also applied their method to coarse-grain
polymeric systems in the spirit described in the previous paragraph. 
The novelty here is the application of the effective potentials 
in coarse-grained simulations by using the dissipative particle 
dynamics thermostat. The results for the present system 
indicate that dynamic properties could be studied using this 
approach. Whether or not this holds true in more complex systems, 
and what are the limits of this approach are topics of further 
studies.

In conclusion, the agreement between the DPD simulations, 
the MD results and experiments strongly indicates that 
the method presented in this work is both physically sound 
and provides a controllable method for performing coarse-graining 
and mesoscale simulations at least to a moderate degree.

\section*{Acknowledgments}

This work has been supported by the Swedish Science 
Council (A.P.L. and A.L.) and the Academy of Finland (M.K.). 
Further support has been obtained from the Academy of Finland 
though its Centre of Excellence Program (I.V.) and from 
the Finnish Academy of Science and Letters (I.V.).

\newpage
\typeout{[Bibliography]}


\begin{thebibliography}{10}
\bibitem{hoogerbrugge92a}
Hoogerbrugge, P.~J.; Koelman, J. M. V.~A. 
{\em Europhys. Lett.} {\bf 1992}, {\em 19}, 155--160.

\bibitem{koelman93a}
Koelman, J. M. V.~A.; Hoogerbrugge, P.~J. 
{\em Europhys. Lett.} {\bf 1993}, {\em 21}, 363--368.

\bibitem{espanol95a}
Espa{\~n}ol, P.; Warren, P. 
{\em Europhys. Lett} {\bf 1995}, {\em 30}, 191--196.

\bibitem{warren98a} Warren, P.~B. {\em Curr. Opin. Coll. Interf. Sci.}
{\bf 1998}, {\em 3}, 620--624. 

\bibitem{flekkoy99a} Flekk{\o}y, E.~G.; Coveney, P.~V.
{\em Phys. Rev. Lett.} {\bf 1999}, {\em 85}, 2522.

\bibitem{flekkoy00a} Flekk{\o}y, E.~G.; Wagner, G.; Feder, J.
{\em Europhys. Lett.} {\bf 2000}, {\em 52}, 271--276.

\bibitem{malevanets99a} Malevanets, A; Kapral, R. {\em J. Chem. Phys.}
{\bf 1999}, {\em 110}, 8605--8613.

\bibitem{malevanets00a} Malevanets, A; Kapral, R. {\em J. Chem. Phys.}
{\bf 2000}, {\em 112}, 7260--7269.

\bibitem{akkermans00a} Akkermans, R.~L.~C; Briels, W.~J. {\em J. Chem. Phys.}
{\bf 2000}, {\em 113}, 6409--6422.

\bibitem{louis00a}
Louis, A.~A.; Bolhuis, P.~G.; Hansen, J.~P.; Meijer, E.~P. 
{\em Phys. Rev. Lett.} {\bf 2000}, {\em 85}, 2522.

\bibitem{bolhuis01a}
Bolhuis, P.~G.; Louis, A.~A.; Hansen, J.~P.; Meijer, E.~P. 
{\em J. Chem. Phys.} {\bf 2001}, {\em 114}, 4296--4311.

\bibitem{Hansen86}
Hansen, J.~P.; McDonald, I.~R. {\em Theory of simple liquids;}
\newblock Academic press: London, 1986.

\bibitem{hend74}
Henderson, R.~L. 
{\em Phys. Lett.} {\bf 1974}, {\em 49A}, 197--198.

\bibitem{murat98a} Murat, M; Kremer, K. {\em J. Chem. Phys.}
{\bf 1998}, {\em 108}, 4340--4348.

\bibitem{reith00a} Reith, D; Meyer, H.; M{\"u}ller-Plathe, F. {\em Macromolecules}
{\bf 2001}, {\em 34}, 2235--2245.

\bibitem{mplathe00}
Baschnagel, J.; Binder, K.; Doruker, P.; Gusev, A.; Hahn, O.; Kremer, K.;
  Mattice, W.~L.; M{\"u}ller-Plathe, F.; Murat, M.; Paul, W.; Santos, S.;
  Suter, U.~W.; Tries, V. 
  {\em Adv. Polym. Sci} {\bf 2000}, {\em 152}, 41--156.

\bibitem{LL95}
Lyubartsev, A.~P.; Laaksonen, A. 
{\em Phys. Rev. E} {\bf 1995}, {\em 52}, 3730--3737.

\bibitem{karttunen02}
Karttunen, M.; Laaksonen, A.; Lyubartsev, A.~P.; Vattulainen, I. 
{\em submitted} {\bf 2002}.

\bibitem{Bes00}
Besold, G.; Vattulainen, I.; Karttunen, M.; Polson, J.~M. 
{\em Phys. Rev. E} {\bf 2000}, {\em 62}, R7611--R7614.

\bibitem{Vattulainen02}
Vattulainen, I.; Karttunen, M.; Besold, G.; Polson, J.~M. 
{\em J. Chem. Phys.} {\bf 2002}, {\em 116}, 3967--3979.

\bibitem{RMC88}
McGreevy, R.~L.; Pusztai, L. 
{\em Mol. Sim.} {\bf 1988}, {\em 1}, 359--367.

\bibitem{reatto86}
Reatto, L.; Levesque, D.; Weis, J.~J. 
{\em Phys. Rev. A} {\bf 1986}, {\em 33},
  3451--3465.

\bibitem{rosenfeld97}
Rosenfeld, Y.; Kahl, G. 
{\em J. Phys.: Cond. Mat.} {\bf 1997}, {\em 9},
  L89--L98.

\bibitem{gonzalez98}
Gonzalez-Mozuelos, P.; Carbajal-Tinoco, M.~D. 
{\em J. Chem. Phys.} {\bf 1998}, {\em 24}, 11074--11084.

\bibitem{LyMar02}
Lyubartsev, A.~P.; Marcelja, S. 
{\em Phys. Rev. E} {\bf 2002}, {\em 65}, 041202.

\bibitem{ostheimer89}
Ostheimer, M.; Bertagnolli, H. 
{\em Mol. Sim.} {\bf 1989}, {\em 3}, 227--233.

\bibitem{soper96}
Soper, A.~K. 
{\em Chem. Phys.} {\bf 1996}, {\em 202}, 295--306. 

\bibitem{toth00}
Toth, G.; Baranyai, A. 
{\em J. Mol. Liquids} {\bf 2000}, {\em 85}, 3--9.

\bibitem{LL_CPL00}
Lyubartsev, A.~P.; Laaksonen, A. 
{\em Chem. Phys. Lett.} {\bf 2000}, {\em 325},
  15--21.

\bibitem{swendsen79}
Swendsen, R.~H. 
{\em Phys. Rev. Lett.} {\bf 1979}, {\em 42}, 859--861.

\bibitem{pawley84}
Pawley, G.~S.; Swendsen, R.~H.; Wallace, D.~J.; Wilson, K.~G. 
{\em Phys. Rev. B} {\bf 1984}, {\em 29}, 4030--4040.

\bibitem{groot97a}
Groot, R.~D.; Warren, P.~B. 
{\em J. Chem. Phys.} {\bf 1997}, {\em 107}, 4423--4435.

\bibitem{For95}
Forrest, B.~M.; Suter, U.~W. 
{\em J. Chem. Phys.} {\bf 1995}, {\em 102}, 7256--7266.

\bibitem{LL97}
Lyubartsev, A.~P.; Laaksonen, A. 
{\em Phys. Rev. E} {\bf 1997}, {\em 55}, 5689--5696.

\bibitem{tourah}
Toukan, K.; Rahman, A. 
{\em Phys. Rev. B} {\bf 1985}, {\em B31}, 2643--2648.

\bibitem{SD94}
Smith, D.~E.; Dang, L.~X. 
{\em J. Chem. Phys.} {\bf 1994}, {\em 100}, 3757--3766.

\bibitem{crc}
Lide, D.~R., Ed. {\em CRC Handbook of Chemistry and Physics;}
\newblock CRC Press: Boca Raton, 82nd ed., 2001.

\bibitem{terama02}
Ter{\"a}m{\"a}, E.; Vattulainen, I.; Patra, M.; Karttunen, M.; Lyubartsev,
  A.~P.; Laaksonen, A. {\em in preparation} {\bf 2002}.

\bibitem{sasha96}
Lyubartsev, A.~P.; Laaksonen, A. 
{\em J. Chem. Phys.} {\bf 1996}, {\em 100}, 16410--16418.

\end{thebibliography}
\end{document}